Title:

HR Del remnant anatomy using 2D spectral data and 3D photoionization shell models

Manoel Moraes[1], Marcos Diaz[2]

1:

Rua do Matão, 1226, Cidade Universitária, São Paulo, SP – Brasil – 05508-090

sala F-308 – mcvmjr@astro.iag.usp.br

2:

Rua do Matão, 1226, Cidade Universitária, São Paulo, SP – Brasil – 05508-090

sala F-303 – marcos@astro.iag.usp.br

Affiliation – 1,2:

IAG-USP: Instituto de Astronomia, Geofísica e Ciências Atmosféricas – Universidade de São Paulo

# ABSTRACT

The HR Del nova remnant was observed with the IFU-GMOS at Gemini North. The spatially resolved spectral data cube was used in the kinematic, morphological and abundance analysis of the ejecta. The line maps show a very clumpy shell with two main symmetric structures. The first one is the outer part of the shell seen in H$\alpha$, that forms two rings projected in the sky plane. These ring structures correspond to a closed hourglass shape, first proposed by Harman and O'Brien (2003). The equatorial emission enhancement is caused by the superimposed hourglass structures in the line of sight. The second structure seen only in the [OIII] and [NII] maps is located along the polar directions inside the hourglass structure. Abundances gradients between the polar caps and equatorial region were not found. However, the outer part of the shell seems to be less abundant in Oxygen and Nitrogen than the inner regions. Detailed 2.5D photoionization modeling of the 3D shell was performed using the mass distribution inferred from the observations and the presence of mass clumps. The resulting model grids are used to constrain the physical properties of the shell as well as the central ionizing source. A sequence of 3D clumpy models including a disk shaped ionization source is able to reproduce the ionization gradients between polar and equatorial regions of the shell. Differences between shell axial ratios in different lines can also be explained by aspherical illumination. A total shell mass of $9 \times 10^{4}$ $M_{sun}$ is derived from these models. We estimate that 50% to 70% of the shell mass is contained in neutral clumps with density contrast up to a factor of 30.

**Key words**: - circumstellar matter - novae, cataclysmic variables -  stars: Kinematics – stars: individual: HR Del – Techniques: spectroscopic

I - Introduction

The classical nova HR Del had its eruption in 1967 as a very slow nova ($t_3$ = 230 days). The spectral evolution of this nova showed nebular emission lines one year after the outburst when the last absorption lines disappeared. In 1975, after its photometric decay, only Balmer lines, [O III]$\lambda\lambda$ 4959,5007, C III / N III blend at $\lambda$ 4640-4650, and He II $\lambda$ 4686 remained detectable.

IUE observations, made between 1979 and 1992, showed that the UV continuum was reduced by 20% and emission line fluxes by 35% (Friedjung et al. 1982 and Selvelli and Friedjung, 2003). The UV luminosity of HR Del at this time was about 56 $L_{sun}$ the highest value found among CVs in quiescence. Observations made by the Einstein satellite in 1980 showed also a bright object in the 0.5-3 keV range (Hutchings, 1980).

An orbital period of 5.14 hours and an orbital inclination of 40(2)° were determined by Bruch, (1982). Kurster & Barwig (1988) suggested that the white dwarf mass is close to 0.6 $M_{Sun}$. The HR Del distance was estimated by several authors and with different techniques suggesting a range between 700 and 1100 pc (Solf 1983, Slavin et al. 1995, Della Valle & Livio 1995, Downes & Duerbeck 2000, Harman and O'Brien 2003).

Longslit spectroscopy observations of the HR Del shell, 15 years after the outburst, made by Solf (1983) showed a highly organized structure. The shell configuration is bipolar with an axial ratio of 1.75 (Harman and O'Brien, 2003). The polar caps have a greater expansion velocity (560(50) km.s$^1$) than the equatorial ring (190(50) km.s$^1$). Solf (1983) suggested a 3D geometrical model for the shell that consists of two polar cones with 50° opening and an equatorial ring at orbital plane. The first published CCD images (Slavin et al. 1994) showed an ellipsoidal object in the light of [O III] and more circular object in the light of H$\alpha$ + [N II]. A similar transition-dependent morphology was seen in early photographs of DQ Her (Slavin et al. 1995). Spatially resolved observations of this nova suggest that a photoionization modeling of its shell should take its peculiar mass distribution into account.

Most of the photoionization models computed so far for nova shells do not reproduce the features seen in the spectra. The simultaneous presence of lines from elements at high and low ionization stages in the spectra and a few inconsistent line ratios are common problems in the models. The interpretation of the observed spectra depends on our ability to reveal the physics, which involves atomic processes that are sensitive to the properties of nebulae. The effects of the geometry and density distribution, like the presence of mass condensations, on the emerging spectra can be very important in some cases, and an unrealistic model may result in an incorrect determination of the nebular and central source parameters as well as biased chemical abundances.

The presence of condensations in nebulae is usually treated in 1D models by adding extra free parameters to an arbitrary radial density profile that reproduces the spectral features. The 1D models have spherical or cylindrical symmetries, which cannot always reproduce the shell geometry. To make a model of the bipolar structures or the presence of condensations we can combine several volume-weighted spherically symmetric models. A pseudo-3D code based on this principle was proposed by Morisset et al. (2005) and has the advantage of being much faster than full 3D codes. Full 3D photoionization codes are based on the Monte Carlo method and can treat the diffuse component of the radiation field self-consistently (for example see the MOCASSIN by Ercolano et al. 2003). The difficulties of full 3D codes are the CPU time demands and the large parameter space to consider for a real nebula calculation, especially when a highly structured environment has to be simulated.

There are about dozen classical novae that have their shell structure spatially resolved and in some of them it is possible to see some structures and asymmetries. HST images of HR Del made by Harman and O'Brien (2003) showed several isolated clumps in the light of H$\alpha$, [N II] and [O III] that have stronger emission than other parts of the shell.

Williams's (1994) analysis of [O I]$\lambda\lambda$ 6300,6364 doublet relative fluxes in several novae showed that the only process that can drive the λ6300,6364 ratio towards unity is the optical thickness in the lines, since high particle and radiation densities drive the emission coefficients to a 3:1 ratio. Considering the O/H abundance in novae, this large optical depth ($\tau \sim 1-2$) cannot be achieved in homogeneous shells. Williams proposed that the [O I] lines come from high density globules in the ejecta, where the gas is essentially neutral.

In the nova shell scenario, the knots or globules seen in the ejecta must originate either early in the outburst or during the evolution of the shell. Anisotropies in early expansion may originate in localized thermonuclear runaways caused by inhomogeneities in the heating (Shara, 1982). In this scenario, part of the shell undergoes rapid expansion because of overpressure, which cools it. The analysis of novae line profiles indicates that individual globules may show heavy element enhancements (Shore et al. 1993). After a period of days the globules may have grown to the size of a normal dwarf star ($10^{11}$ cm) and cooled to the point where their cores are neutral (Williams 1994). There are no studies about the photo-evaporation of these globules by the central source radiation in novae shells. In the planetary nebulae (PNe) scenario, calculations made by Mellema et al. (1998) showed that clumps can survive 250 years before half of their mass evaporates and their life time is near 1000 years. But in their calculations the clumps were bigger ($2 \times 10^{16}$ cm), more massive ($2.3 \times 10^{5} M_{sun}$) and more distant from the source of radiation ($10^{18}$ cm) than they are in a novae shell environment. Bertoldi & McKee (1990) deduced that the pressure of photo-evaporation is proportional to the square root of the ratio between the ionizing flux and the product of the distance to the central source and the size of the clump (their equation 3.23). If a simple scaling for novae conditions is made, the photo-evaporation pressure obtained through this may be 40 times bigger than the typical PNe values. As the mass loss rate of clumps is proportional to the photo-evaporation pressure, the life time of clumps in novae could be much shorter. However, the radiation flux on the clumps depends on the absorption by hydrogen along the photon's path from the central source to the clump as well as the interaction and occultation between clumps. The occultation from the radiation source may eventually increase the survival time by 30 % (Lim & Mellema 2003). Images of the nova remnant in GK Per show clumps 100 years after outburst (Slavin et al. 1995 and Bode, 2004). A more detailed calculation is still needed in order to provide realistic life times for clumps in novae shells.

In this paper the spatially resolved emission line spectra around nova HR Del obtained by GMOS/IFU are presented and the first application of a pseudo-3D code, named RAINY3D, (Diaz 2001) is described, which uses CLOUDY (Ferland, 2005) to calculate photoionization and the radiation transfer in a 3D novae shell model with condensations. This code is applied to the spatially resolved spectral observations of HR Del to determine the parameters of the central source and ejecta.

II - Observations

The observations were made using the Gemini Multi-Object Spectrograph (GMOS) and its Integral Field Unit (IFU) at Gemini Telescope North. The spectral range was from 0.4 μ to 1.0 μ and with a spatial resolution of 0.5". The exposure time was set to 600 seconds for each frame. To cover HR Del envelope completely, a mosaic of 12 frames was used, covering 13"x14" of the sky. Each frame contains 500 microlenses of 0.20"x0.20" each and covers 5.0" x 3.5" in the sky. The sky background was sampled with 250 microlenses (5.00" x 1.75") which were located at about 1 arcmin away from the object. The seeing in the V band during observations was close to 0.5". The observations were taken at an airmass close to 1 in order to reduce undesirable atmospheric dispersion effects (1.00 to 1.16). Sky transparency conditions during all nights allowed spectrophotometric calibration.

We used the R150+G5306 grating in single mode centered at 5070A, giving a sampling of 0.17 nm per pixel which means ~ 70 km.s$^{-1}$ per pixel at Hα. The instrumental FWHM profile is about 300 km.s$^{-1}$.

The data reduction was made using IRAF v.12.2 routines. The spectra were sky subtracted and corrected for bias. Flat field frames were used to normalize the response of each microlens. Since each lens aperture gives one spectrum, we have obtained a total of 6000 spectra in the HR Del mosaic. Variance weighting extraction of individual spectra was performed and the wavelength calibration was done using Cu-Ar arc lamp images. The sensitivity function derived from standard stars was applied to all the microlenses in the observed field. In the nebular region, the typical continuum signal to noise ratio (S/N) in one resolution element is ~10 near the H$\alpha$ line.

The individual frames were manipulated with IRAF tasks and custom FORTRAN programs in order to produce a data cube that contains all the spectral and spatial information. The mean value was used for the lines fluxes measured in overlapping spatial regions. The original IFU spatial resolution of 0.2" was maintained but a Gaussian convolution was used on the emission line maps to control the trade-off between S/N and spatial resolution. The local continuum was subtracted from each frame to generate the emission lines maps.

III – Spectral imaging of the resolved shell

The GMOS-IFU data cube can be used to generate emission line maps of HR Del's shell at any given velocity. Figure (1) shows the maps of H$\alpha$ + [N II] and [O III]$\lambda$5007 lines. We can see that the shell has a bipolar symmetry with polar caps and an equatorial ring. The axial ratios obtained from each map are different, from H$\alpha$ + [N II] map is 1.60 while for [O III]$\lambda$5007 is 1.95. This result is in agreement with the observations by Slavin et al. (1995) who obtained 1.5 and 2, respectively. To perform this calculation, the orbital inclination value of 38° (Solf, 1983) was assumed. This morphological effect was also observed in DQ Her (Baade 1940). The axial ratio seems to be related to the speed class of the nova, where higher values of axial ratio are seen in slower novae (Slavin et al. 1995). Therefore, HR Del fits within this relation.

The images in figure (1) show a prolate remnant with dimensions of 16 x 10 arcsec in H$\alpha$ + [N II] and 13.7 x 7.2 arcsec in [O III]$\lambda$5007, with these dimensions representing the extended nebular emission above 1 sigma background noise. The H$\alpha$ + [N II] shell shows a diffuse component which is larger than [O III] by 17%. A line map with wavelength centered in [N II]6584 also shows a smaller shell, by 13%, when compared to the H$\alpha$ + [N II] shell. The major axis displays a position angle of 45° and the minor axis is believed to be close to the orbital plane. Each polar cap has a 90 degree cone from the center.

The expansion velocity of the polar and equatorial components of the shell can be obtained from our data. The polar components have a velocity of 560(60) km.s$^{-1}$, as calculated from the [O III] $\lambda\lambda$4959,5007 lines, while the equatorial components have a velocity of 300(60) km.s$^{-1}$. These values are in agreement with the axial ratio and also with the velocity obtained by Harman & O'Brien (2003) of 320 km.s$^{-1}$. A distance through the expansion parallax of 850(90) pc is estimated, which is in broad agreement with the values found in literature (700 - 1100 pc) derived by several methods and authors.

The [O III] and H$\alpha$+[N II] maps show that the shell is very clumpy and most of the emission comes from the polar structures. The emission lines formed in the brightest parts of the polar structures are, on average, ten times brighter than in the dimmer parts of the shell. These features occur on both line maps and could be explained by the matter distribution, where the denser regions have a larger line flux. The H$\alpha$+ [N II] map shows a region of high emission at the edges of the equatorial axis, and this structure cannot be seen on the [O III] map. This could be due to the abundance or ionization differences with respect to those in the polar regions. The presence of an accretion disk at the orbital plane could absorb some ionizing photons from the inner part of the disk as well as from the boundary layer and this could reduce the O$^{++}$ population in the equatorial region. The H$\alpha$+ [N II] line map shows a shell bigger than the one seen in [O III]. This effect is expected to be observed and is caused by a lower ionizing flux at the outer regions of the shell. On both line maps the top polar structure, which looks like a parachute, shows a different shape when compared to the bottom polar cap, but these structures are equidistant from ionizing source. In figure 4 the line ratio is seen between H$\alpha$+ [N II] and [O III]$\lambda$5007 providing a clear view of the shape of the polar caps, where the [O III] emission is strong. The external regions have enhanced H$\alpha$+ [N II] emission and it appears as an elliptical ring around the polar structures. This was also observed on the HST + WFPC2 images published by Harman and O'Brien (2003). The total flux in H$\alpha$+ [NII] produced by polar structures accounts for 68% of the total line flux, but the [O III] lines in the same structure correspond to 91% of the total [O III] emission.

At least two clumps of enhanced emission in H$\alpha$+ [N II] line (figure 1) can be seen outside the shell. The bottom one is also seen on the [O III] map, but the top one is barely seen on this map. It is possible to ask if these clumps belong to the shell or are an instrumental artifact. The first thing to consider is that these clumps can be seen in other strong lines. Second, the clump closer to the equatorial axis does not show strong [O III] emission, following the behavior of the equatorial region while the clump next to the polar axis shows strong [O III] emission, like the surrounding regions. This suggests that the clumps are close to the shell. Third, the clump next to the polar axis shows lines that are redshifted, like the lines of the bottom polar cap (figures 2 and 3) and cannot be seen on the blueshifted line map. Fourth, Harman and O'Brien (2003) showed that in 1997 there were knots with a velocity higher than the shell, but they could not determine if these features were located outside the shell. Our observations suggest that these clumps are part of the ejected shell and are located outside it.

Since the shell has a high expansion velocity, the lines are substantially redshifted at the bottom polar cap and blue shifted at the top polar cap. At the H$\alpha$ wavelength this cause a blend between this line and [N II]$\lambda\lambda$ 6548,6584 lines and this will certainly affect the measured Balmer H$\alpha$/H$\beta$ ratio. The velocity structure also makes some line identifications difficult. Emission line maps of [O III]$\lambda$5007 are shown in intervals of 120 km.s$^{-1}$ and over a velocity range from – 900 to + 540 km.s$^{-1}$ in figure (2), presenting the complex velocity structure in this line. It is also possible to see that the shell is very inhomogeneous and there are many clumps with different velocities. Figure (2) also shows that close to the equatorial regions the emission is almost at rest. This behavior is also seen on the H$\alpha$+ [N II] line maps shown in figure (3). These maps display some amazing symmetric structures and also several differences when compared to the [O III] line maps of figure (2). There are two eccentric clumpy rings on the maps in the range of 6555-6560 A (figure 3F) and 6565-6570 A (figure 3H). The rings have about the same size and thickness. Some clumps have a size comparable to the ring thickness, but it is possible to see even smaller structures in the shell. The structure of these rings is resolved for the first time in a classical novae shell remnant. A bipolar shell resembling a closed hourglass was first proposed by Harman and O'Brien's (2003) kinematic model when they observed HR Del with HST. The rings have a diameter of 8.0 arc seconds and a 1.7 arcsec thickness (or 6.45 x 10$^{16}$ cm and 1.31 x 10$^{16}$ cm at 850 pc). It is also possible to see these ring structures on the panel of 6585-6590A (figure 3L) close to the rest wavelength emission of [N II] $\lambda$ 6584. The map in the range of 6540-6545A (figure 3C) also shows a ring close to the wavelength of [N II] $\lambda$6548. These eccentric rings are certainly farther away from the central source than the polar caps structures seen on the [O III] maps. The clumps located at the equatorial regions seen on the 6560-6565A map (figure 3G) can be identified as the overlap of the two rings.

The three maps in the range of 6530-6545A (figures 3A, 3B and 3C) show a blue shifted polar cap and the three maps in the range of 6590-6605A (figures 3M, 3N and 3O) display the red shifted polar cap. Both show the same clumpy structure that appears on the [O III] $\lambda$ 5007 maps (figure 2). On the maps showing the blue shifted polar caps (figure 3A, 3B and 3C), the clumps seem to be the emission of [N II]$\lambda$6548 while on the maps of the red shifted polar cap (figures 3L to 3O), the emission could be from [N II]$\lambda$6584. The red shifted [N II]$\lambda$ 6548 emission appears in the range of 6555-6560A, (figure 3F) and the blue shifted [NII]$\lambda$6584 appears in the range of 6570-6580A (figures 3I and 3J). The expansion velocity was measured by using the H$\alpha$ lines from figures 3FGH and obtained 630(60) km.s$^{-1}$. This value is in agreement with the expansion velocity obtained from the [O III] lines of polar structures when we consider the uncertainty of the determination. But if we consider the absolute value of each determination, the distance of the polar caps structures from the external ring should be ~1 arcsec in radial direction at the observation time. This is the approximate distance measured on the line maps of the ejecta.

IV - Line Identifications and overall shell diagnostics

Since the shell has a complex velocity field the line identification was done locally. In order to obtain the nebula lines fluxes we divided the maps in two regions. The first one is centered on the central source and has a 2 arc seconds radius. The second is the remaining region, centered on the central source, which has a 2-arc-second inner radius. All line intensities were normalized to the total H$\beta$ emission flux in the outer region (1.19x10$^{11}$ erg/s cm$^2$) and were dereddened with E(B-V) = 0.16 (Selvelli and Friedjung, 2003). Table 1 shows the proposed line identifications and their flux relative to H. Other lines above 8000 A like the HI Paschen lines were observed, but unfortunately their fluxes were contaminated by second order spectra. The central region shows a relatively small contribution from the nebulae, since the shell surrounds the bright accretion disk in the binary. There are lines of high ionization species, like Fe VII, Ca VII and C IV and also lines from neutral atoms and lower ionization stages, like H I, He I, O I and N II. The permitted lines have most of their flux coming from the region closer to the central source, like H$\gamma$, C III+N III and He I 6678, which have 90% of their flux from this region. Other lines, like [O III]$\lambda\lambda$4959,5007, and [O I]$\lambda$6300 are formed in the shell.

Some line ratios expected from atomic theory for an optical thin gas were not observed. The intensity ratio of the [O I]$\lambda\lambda$6300,6364 doublet is expected to be 3:1, for an optically thin gas, but the observed ratio is 2:1 or even 1.2:1 in some regions. Williams (1994) observed that, particularly for Fe II novae, these lines usually appear in the shell evolution and their strength is difficult to explain. He suggested that the shell ejecta is not homogeneous and that the [O I] emission comes from clumps, where neutral gas could exist. It is amazing that HR Del has [O I] lines 12000 days after the eruption and this could be an evidence that neutral clumps can survive for a long time after the eruption. Using the [O I] $\lambda\lambda$6300,5577, [O I]$\lambda\lambda$6300,6364 ratios and the expression given by Williams (1994) one can obtain the optical depth of the [O I] line and estimate the electronic temperature where the line is produced. A T$_e$ = 6500(500)K was obtained for $\tau_{6300}$ = 2.6. The temperature uncertainty arises from flux measurement errors. The observed [O III] 4959,5007 ratio is close to the ratio between transition probabilities. The H$\alpha$/H$\beta$ line ratio was expected to be close to the recombination value, but the H$\alpha$ line, because of expansion, is blended with the strong forbidden lines of [N II]. The observed (H$\alpha$ + [N II]) / H$\beta$ line ratios have a range from 2.9 in the central regions to 10 in some polar cap regions.

The observed He I 5876 / H$\beta$ intensity ratio is between 0.1 and 0.2 with the lower value found in the central source region while the shell has a ratio close to 0.2. If we consider case B recombination, then the observed He I 5876 / He I 7065 ratio of 15 would correspond to a temperature close to 5000K (for N$_e$ = 10$^4$ cm$^{-3}$). An increase in Arial the shell density would diminish the ratio and the same would cause an increase in the line optical depth (Almog and Hetzer, 1989). Clegg (1987) showed that the He I 7065 line has a collisional to recombination excitation rate 10 times larger than the value of He I 5876 and this means that the He I$\lambda$7065 is more affected by collisional excitation. So the high observed ratio may be caused by a low gas temperature. Other Helium line ratios (He I 5876/He I 6678 and He I 6678/HeI 4922) are far from the expected case B values. The He I 5876 / He I 6678 ratio has a small variation with T$_e$ and it is expected to be close to 4, but it was found to be 9.6. We could not explain this large value by a blended line next to He I 5876. Another effect to be considered is collisional excitation which may increase the emissivity of He I 5876 two times more than the increase of He I 6678 (a singlet line) for the same temperature, (Ferland, 1986). This could mean that the shell is indeed clumpy.

We estimate the abundance of He relative to H in the shell assuming an electron density of $10^3$ cm$^{-3}$ and a temperature of 7500 K. The values of recombination coefficients for He were interpolated from Osterbrock (2006) for 5000 K and 10000 K. The effects of collisional excitation from the metastable 2s $^3$S level of He I 5876 were corrected using the factor (C/R=0.03) given by Clegg (1987). The He lines used in the calculation were He I 5876 and He II 4686, yielding nHe / H = 0.21. Tylenda (1978) obtained nHe / H = 0.17 from the single blackbody model and 0.27 from two blackbody model. This value is not very sensitive to assumed electron temperature and density. In the shell we have a nHe I / He II ratio of 1.77, or 64% of neutral Helium.

The Zanstra temperature for the central source was derived from the He II$\lambda$4686/H$\beta$ line flux and non-LTE model atmospheres by Clegg and Middlemass (1987). For log(g) = 7.5 and Y = 0.1, a T$_{Zanstra}$ = 85000 K was found.

V – Shell analysis

V.1 - Runaway clumps

As said above, isolated clumps on several emission line maps (figure 2 and 3) were detected. At the left side of the shell there are two clumps that seem to be at the edge of the shell or perhaps outside the shell. They can be seen in redshift (figures 3F, 3G and 3H) and in blueshift (figures 3L, 3M and 3N). The one closer to the polar axis has red shifted emission lines (500 km.s$^{-1}$), but because of projection effects, whether it is a high velocity clump cannot be said. This clump appears to rise at 6550-6555A, figure 3-E and fade at 6565-6570A (fig. 3H). It rises again at 6585-6600, (figures. 3K, 3L, 3M), due to the redshifted [N II] $\lambda$6584 emission and it is also seen on the red shifted [O III] line maps. The clump close to the equatorial regions has a low [O III] / H$\beta$ emission line ratio when compared to the other clump. This clump appears at 6545A (fig. 3D) and fades down in the range of 6570-6575A, (fig. 3I). It rises again in the range of 6580-6585A, (fig. 3K) and disappears in more redshifted wavelengths. Both clumps show high [O I] emission, which corresponds to 25% of the H$\beta$ emission, compared to only 5% in the polar caps regions. High values of O I / O II (five times) and O II / O III (four times) ratios were found in the clumps, when compared to other regions of the shell. These line ratios show that the conditions inside the clumps are better for lower ionization or neutral emission to occur, as suggested by Williams (1994). The Balmer H$\alpha$/ H$\beta$ ratio is between 5 and 6 which means an important contribution of [N II] lines, since we expect values close to 3. Only in these clumps it was possible to measure the [N II] $\lambda$5755 emission line.

There is a clump at the top of the shell (figure 3) in the range of 6560-6580A (figures 3G to 3J) that appears to be in the shell on the map 6560-6565A (fig. 3G) and outside the shell in the range of 6570-6580A (figures 3I and 3J). These clumps show similar line ratios of other clumps. At the bottom of the red shifted polar cap, in the range of 6580-6585A (fig. 3K), there is another clump at the edge of the shell.

V.2 – Spatially resolved spectroscopic analysis

The HR Del shell has a very heterogeneous matter distribution as discussed above. In order to study abundance gradients and other physical conditions variations along the shell, the spatially resolved shell was first divided into regions along the polar and equatorial axes as shown in figure 5. For each region the average spectrum was obtained and used in the line ratio analysis. The regions had their projected distance from the central source roughly estimated by assuming an hourglass geometry. However, there are projection effects that could not be taken into account, and there may be some contamination by regions that are farther or closer than the estimated value. Those along the polar axis were also corrected from projection effects using the orbital inclination angle. The emission line analysis is important to determine if the shell abundances and/or ionization conditions vary along the shell. Farther regions receive a lower ionizing flux, since the number of ionizing photons decay at least with $r^{-2}$. With the goal of obtaining the local He abundances, the same temperature and electron density were used for all regions. Figure 6 shows the $He^+$ ion abundance relative to Hydrogen in polar and equatorial axis. The positive values of axis represent the blue shifted polar cap and the south (bottom) part of the shell. We see that the $He^{++}$ population, which, through recombination produces the He II lines, decay in a similar way in both axis and in the both sides of the shell. The decay scales roughly as $r^{-2}$ from the center to $2 \times 10^{16}$ cm in both directions and for both axes, as expected from the dilution of ionizing photons. Another effect to consider is that along the path from the central source to the outer parts of the shell, the most energetic photons that ionize $He^+$ are removed from the radiation field. It is possible to notice that in the equatorial axes the decay is a bit faster than it is along the polar axes. This could mean that the shielding produced by the accretion disk affects the ionization in equatorial regions. For regions that are farther than $2 \times 10^{16}$ cm it is possible to see that the $He^+$ population relative to H is approximately constant until the outer edge of the shell. This occurs because the He density, like the ionizing flux, also decays with distance while the ionized fraction remains nearly constant. The He ion abundance profile is showed in figure 7. One can see that the He II and He I abundances have a very similar profile in both axes. In the equatorial axis the total He abundance remain close to a constant value of ~0.2. In the polar axis there are some fluctuations near the polar caps, where the estimated He abundance by number is 0.15 at both sides of the shell. These regions correspond to the clumpiest part of the shell. The He I lines were corrected from collisional effects (C/R correction) with a mean value for all regions of the shell. Since the clumpy regions are denser than the mean shell, the C/R corrections are greater at the clumps. The results suggest that the collisional effect is not affecting the He I abundance estimates at the clumps. The temperature affects the abundance determinations by the recombination coefficients of H$\beta$, He II $\lambda$4686 and He I $\lambda$5876 lines. For the He I line, the ratio of recombination coefficients used to calculate the $He^+/H^+$ abundance remain nearly constant (within 3%) from 10000 to 5000 K. For the He II line, the ratio of H$\beta$ and He II $\lambda$4686 recombination coefficients used to calculate $He^{++}/H^+$ abundance decrease only 10% from 10000 to 5000 K. The conclusion is that the temperature determination errors will not produce the abundance fluctuations observed in the clumpy polar axis regions. The flux measurement errors also affect the abundance determination. They were obtained from regions which are composed by at least 100 micro lenses with combined spectral S/N > 20. Both He II $\lambda$4686 and He I $\lambda$5876 are strong lines and there are no other strong emission lines observed in novae remnants in these wavelengths, so the measurement flux error is determined by the uncertainty in the continuum level, which is close to 5%. The conclusion is that this fluctuation may be possibly caused by the presence of neutral Helium in the clumpy regions.

In order to obtain the oxygen abundance, the classical nebular line analysis was performed to estimate the $O^{++}$, $O^+$, and neutral O population from the forbidden line fluxes. The task IONIC/IRAF was employed to solve an 8-level-atom matrix and compute level populations and abundance for a single ion for a given electron temperature and density (De Robertis et al. 1987 and Shaw & Dufour, 1994). The intensities relative to H$\beta$ of [O III]$\lambda$5007, [O II]$\lambda\lambda$7320,7330 and [O I]$\lambda$6300 lines was measured to obtain the ion abundances relative to Hydrogen. The H$\beta$ emissivities derived using the formula by Aller (1994) are accurate to within 4%, for electron densities fewer than $10^6$ cm$^3$. The relative Oxygen to Hydrogen ions abundances depend on the local electronic densities. Since the HR Del shell has a very heterogeneous density profile, it is necessary to account this feature in the abundances determinations. For the typical densities of the old novae remnants, only the OII abundance determination has a density dependence, because the [O II]$\lambda$7325 critical density is close to the mean shell density values. For the same temperature and Oxygen line intensities, a lower density environment results in a larger $O^+$ abundance value. To perform the abundance calculation it was assumed that the density of each region is defined by the best clumpy model geometry and density, (see section V). The clumpy regions have a mean density of 2000 cm$^3$ and the external regions 300 cm$^3$.

Figure 8 shows the oxygen abundance across the polar and equatorial axes. The shell spectra present some O III recombination lines, such as $\lambda 5592$ and $\lambda 7713$ which indicate that there are $O^{+3}$ ions in regions closer to the central source. The outer edge of the $He^{++}$ zone at $2 \times 10^{16}$ cm is also the outer zone of $O^{+3}$ and the inner zone of $O^{++}$, since the $O^{++}$ ionization potential is 54.9 eV, nearly the same as $He^+$. Therefore, the oxygen abundance gradient from the central regions of the shell to the clumpy regions seen in figure 8 may be caused because the $O^{+3}$ contribution to the total oxygen abundance was not considered.

Figure 9 shows the oxygen ionization profile along the polar and equatorial axes. In the equatorial axis the $O^+$ population (90%) is much larger than the $O^{++}$, but these ions have similar abundances (50%) along the polar axis. This indicates that in the equatorial regions are less ionized than the polar regions (figure 6). The abundance profile in the equatorial and polar axes has a similar behavior at the same distance from the ionizing source. More [O I] emission is expected to be seen in the outer regions of polar axis because there is a decay in the $O^+$ and $O^{++}$ abundances, but it cannot be measured. This could be caused by the low electron temperature in these regions, below the excitation temperature of [O I] transitions. The [O III]$\lambda 4959$ transition has a smaller critical density for collisional de-excitation than [O III]$\lambda 5007$. Because of this effect, this line ratio was found to be greater than expected from theoretical transition probabilities for denser regions. For a nebula with 10,000 K the critical density of [O III]$\lambda 4959$ is 510 $cm^3$ while it is 3600 $cm^3$ for [O III]$\lambda 5007$ (Osterbrock 2006). These values can be regarded as a lower limit for the clump density where the [O III] line ratio was greater than expected.

Hummer and Stoney (1987) calculated the Hydrogen line ratios for a large range of temperatures and densities. For the physical conditions found in novae ejecta, which means electronic temperatures from 1000 K to 12500 K and densities from 100 $cm^3$ to $10^6$ $cm^3$, the $H\alpha/H\beta$ theoretical ratio is in the range of 2.8 to 3.7, where the larger values are found in lower density and cooler environments. The effect of ionization conditions in the behavior of the line ratio along the axis can be seen in figure 10. In clumpy regions the [N II] lines contribute more to the $H\alpha$ + [N II] blend flux and a larger line ratio to $H\beta$ can be seen. The contribution of [N II] is large and in some regions at the polar caps the ratio is close to 10. The polar regions have a $H\alpha$ + [N II]/$H\beta$ ratio between 6 and 10, while the equatorial regions have a ratio between 3.7 and 4.2. Therefore the [N II] $\lambda\lambda 6548,6584$ emission lines have the same spatial distribution behavior of [OIII] lines in agreement with an aspherical illumination. However, there is also a decrease of the Nitrogen line ratios in regions farther than $4 \times 10^{16}$ cm for the polar axis and farther than $3 \times 10^{16}$ cm in the equatorial axis. This could be due to the decay of $N^+$ population and we should expect to see some N I lines in these regions, like $\lambda\lambda$ 7442,7468, which are the strongest lines of N I in the observed wavelength range, but no N I line was observed. One possibility to explain this feature seen in oxygen and nitrogen lines is the lower abundance of these atoms in the external regions of the nebula. By looking at the $H\beta$ line map (figure 11) we find no enhanced emission at the polar caps regions, in contrast with the $H\alpha$ + [N II] blend map. This corroborates the idea that the polar caps regions may have different Nitrogen abundance when compared to the external polar regions of the shell. In order to maintain a constant Nitrogen abundance through all the polar axes, the N I $\lambda 7468$ emission at outermost regions would have to be ~10% of $H\beta$.

V.3 – Photoionization models

Three different types of photoionization models were computed for the HR Del ejecta. The first one is a classical model for an homogeneous shell without condensations. This model shell has a density profile following a power law $r^\alpha$ and has its dimension estimated by the velocity field measured in the HR Del shell. The second shell model was constrained by the IFU-GMOS observations discussed in sections III and IV. The model shell has an hourglass shape with geometry constrained from the emission line maps. The third model uses the same shell configuration of the second model, but the effects of the central source geometry and condensations were explored.

The input lines were used as given in table 2 for all models calculations and spectral analysis. Some upper limits were obtained from the HR Del data and were also used to constrain the models. The line fluxes were then compared to the observed shell fluxes (table 1). To model only the ejecta of HR Del, the contribution from the central source were eliminated by removing the microlenses that were up to 2 arc seconds from the stellar image center. The nebular spectra of HR Del were corrected from reddening using $E_{(B-V)} = 0.16(2)$ (Selvelli and Friedjung, 2003). Since the central object is very bright, it is not possible to rule out some contamination by the central source spectra in the innermost nebula microlenses. Another effect to be considered is the intrinsic scattering in the nebula, but it is not possible to measure this process from the data at hand.

V.3.1 – Basic parameters and homogeneous models

Simple spherically symmetric shell models were computed to explore the physical conditions in the nebula. All calculations were performed using CLOUDY 6.02 (Ferland 2005). The central source parameters were constrained by IUE spectral observations of HR Del made between 1979 and 1992 (Selvelli and Friedjung 2003). These authors observed that the UV continuum decayed by a factor of 0.8 while the line intensities decayed 35 % (He II $\lambda$1640) to 55% (C IV $\lambda$1549) in the 1979-1992 period. These line intensities, scaled to the HR Del GMOS observation time, were adopted as upper limits for the model line optimization. IUE data also give us upper limits for the central source luminosity and temperature for a distance of 850 pc. The ionizing radiation field was assumed to be a high gravity hot star atmosphere. Ionizing spectra from Rauch (2003) hot stellar atmospheres with log(g) = 7-8 and temperatures from 50000K to 150000K were used in the calculations.

It also possible to roughly estimate the mass accreted by the white dwarf before the eruption. The analytic work by Shara (1981) showed that the critical pressure at the base of the envelope to trigger a TNR depends on the white dwarf luminosity by the power of 5/8, the WD mass and the mass accretion rate. Simple lower limit values to the envelope mass may found using this prescription. However, the ejected mass during outburst is also related to $t_3$ and the shell expansion velocity, that are observable parameters for novae. Detailed simulations were made by Yaron et al. (2005) varying the values of three basic independent parameters: the WD mass, WD temperature, which is related to its luminosity, and the mass transfer rate. For the $0,67(8)$ $M_{sun}$ white dwarf of HR Del (Ritter and Kolb 2003), a decay time $t_3$ of 230 days and expansion velocities of 550 km.s$^{-1}$ observed in the HR Del remnant, their best model gives an envelope mass of $\sim 1 \times 10^{-4}$ $M_{sun}$. Their simulations showed also that a high ejected mass occur on less massive and/or lower luminosity WDs, and for mass transfer rates close to $10^{-10}$ $M_{sun}.yr^{-1}$.

On the other hand, it is possible to estimate the ejected shell mass by the H$\beta$ luminosity, since it depends on the H$\beta$ emissivity, the shell temperature, the mean molecular weight and the electron density, again adopting a distance of 850 pc. Using $T_e = 6000$K and $N_e = 1000$ cm$^{-3}$ (mean values for the shell) one finds $1.24 \times 10^{-3}$ $M_{sun}$ for the total shell mass, which is greater than the accreted envelope mass needed to trigger an eruption. This value is regarded as an upper limit, because some contribution of the central object to the H$\beta$ intensity of the shell may be present in the inner nebula spectra. The neutral oxygen mass is also estimated by using the relation given by Williams (1994) and the result obtained was that it is only 0.01% of the total oxygen mass. The mean oxygen mass fraction of novae is about 15% of the total shell mass (Gehrz et al. 1993). Considering this value, the neutral oxygen mass obtained for novae at LMC showed that the masses of the ejecta are substantially greater than previously believed, exceeding values of $3 \times 10^{-4}$ $M_{sun}$ (Williams, 1994). Therefore, the neutral oxygen mass also indicates a high ejecta mass for HR Del.

Some parameters of the shell that were obtained from the previous spectral analysis, like the He abundance (see sections III and IV), were kept constant through all the calculations to avoid a large number of free parameters. The adopted Oxygen abundance is just a lower limit, since only the abundance of the ions $O^{++}$, $O^{+}$ and $O^{0}$ was obtained and did not account for the $O^{+++}$ contribution. The inner parts of the shell have OIII recombination lines such $\lambda$ 5592 and $\lambda$ 7713, therefore there are $O^{+++}$ ions. However, if we consider that in the clumps the $O^{+++}$ contribution is much smaller than the other ions, the abundance value for the clumpy regions should be close to the total oxygen abundance. In the regions that are clumpy we found log(O/H) = -2.52. This value is close to Tylenda (1978) calculations, who obtained log(O/H) = -2.35 and Sanyal (1972), log(O/H) = -2.4. The Nitrogen abundance adopted is log(N/H) = -2.5, as obtained by Tylenda (1978). The carbon abundance used here was also determined by Tylenda observations of HR Del, as log(C/H) = -3.4. These abundances were assumed as initial values and additional models with variable CNO abundances were computed.

Models luminosities around $\log(L_*) = 36$ erg.s$^{-1}$ were obtained, constrained by the UV flux observed by IUE. All symmetric models fail to reproduce the observed line ratios [O III] $\lambda$5007/H$\beta$, He II $\lambda$4686/H$\beta$ and He I $\lambda$5876 / H$\beta$ simultaneously when the Oxygen abundances measured in the clumps are used. The model with central source temperature of 61000 K has an electron density, $N_e$, ranging from 1500 to 400 cm$^3$, and a shell mass of 1.75 x 10$^{-3}$ $M_{sun}$ and it is able to reproduce the [O III] / H$\beta$ and HeI/H$\beta$ line ratios, but it could not match the observed He II 4686 / H$\beta$ ratio. The electron temperature of the model shell is close to 6000K, which is in agreement with the temperature estimated by the [O III] line ratio.

The model could not also reproduce the low ionization ions lines, such as [O II] $\lambda$ 7325 blend and [O I] $\lambda$6300. The [O II] line intensity in the HR Del shell is 2 orders of magnitudes larger than that predicted by the model without clumps. For the [O I] lines the discrepancy is even larger.

This simple model without clumps may be considered as the zero order model and gives us elements to compare the following 3D models with condensations as well as with models with a disk-type central source.

V.3.2 - Models with condensations

The line maps shown in section III display a very structured shell with many regions where the matter is condensed into clumps. The shell spectra have [O I]$\lambda\lambda$6300,6364 lines and their ratio still indicates a high line optical depth between 1 and 2. Unfortunately, these lines are too noisy to produce a line map that could show where in the shell the denser clumps are. The [O III]$\lambda\lambda$5007,4959 ratio is close to the transition probabilities ratio of 3:1. Therefore, there is the coexistence of optically thick [O I] emission and optically thin [O III] emission which indicates the presence of dense clumps in the shell.

To compute a realistic photoionization model of an ejecta with condensations a 3D description of the shell is necessary. In this section, the methods used to perform a photoionization model of a 3-D shell with condensations are described. CLOUDY was used to compute the radiation-gas interaction and follows all relevant physical processes and quantities of a NLTE nebula, including the ions and levels population, the electron temperature and local line intensities. A code named RAINY3D proposed by Diaz (2001) was used to build a pseudo 3D shell with condensations, driving CLOUDY as a subroutine. To perform the HR Del ejecta model GMOS and observational data from literature to constrain the shell geometry and the central source parameters.

The shell model has two density components, the first is the background and the second is the clumps. The material that is not in the clumps belongs to a background, which has a radial density profile like $r^{-\alpha}$. The shell is characterized by its condensed mass fraction, (fc), which is the ratio between the mass in the clumps and total shell mass. This shell has its dimensions determined by the velocity field measured in the HR Del spectra, where the maximum velocity gives the outer radius ($r_{out}$) and the minimum velocity gives the inner radius ($r_{in}$). The $\alpha$ value was obtained by measuring the H$\alpha$ intensity in regions between $r_{in}$ and $r_{out}$ and a radial density profile was fitted using a power law, with the best fit $\alpha = 0.9$.

The clumpy component was determined by the emission line maps. The clumps were built assuming a Gaussian profile, where the clump size was defined as 2*FWHM and their maximum density values were normalized by the local background density. We assumed a 40° shell inclination and an hourglass geometry (see section III and IV). The velocity field maps seen in figures 2 and 3 were used to estimate the condensation dimensions, densities and their position in the shell (assuming the hourglass geometrical structure).

The code RAINY3D was used to build a special grid to and interact with the main photoionization code, CLOUDY. To build a 3D shell with CLOUDY which is a 1D code, it is necessary to generate radial profiles for each direction from the central source to the shell. The solid angle of each 1D calculation is defined by an adaptive angular grid, which depends on the minimum clump size and the maximum radius of the shell. If we have a large shell with small clumps, the angular step will be small and the number of 1D calculations will be very large. CLOUDY computes the emitted fluxes for each direction and the sum of all, weighted by the solid angle of each beam, corresponds to the total shell flux. Table 3 shows the input parameters of the all models.

Two ring-like structures which define the outer parts of the ejecta and two polar cap structures inside the rings were identified. Both structures are very clumpy, but we identified that the clumps in ring structures have a size comparable to the ring thickness, and the clumps seen in the polar cap regions look to be smaller. The clumps located in the polar regions have their size measurements limited by the instrumental spatial resolution. The clumps size is less than 0.5 arcsec or $4.8 \times 10^{15}$ cm. They can be seen in figures 2 and 3. From their Balmer emissivity, it was found that they have a maximum density 10 times larger than the surrounding background. A deconvolution by a Gaussian instrumental PSF suggests that the maximum density of these clumps could be 16 times larger than the background. The overall shell structure was divided in two main parts. The first one reproduces the rings structures seen on H$\alpha$ line maps (figure 3) which are at the outer part of the shell. In the 3-D model this structure is composed by two spheres merged in an hourglass shape. The sphere's diameter is 8.0 arcsec or $6.45 \times 10^{16}$ cm at 850 pc, and thickness of 1.7 arcsec or $1.3 \times 10^{16}$ cm. The space inside the spheres is filled with diffuse gas that has a power law density profile running as $r^{-\alpha}$ and is named background component. The second structure reproduces the globules seen on the O III line maps (figure 2). These shell features are located inside the hourglass structure.

The oxygen line maps show that most of the O III emission comes from the polar regions what could mean that the gas is denser in the polar caps. But the oxygen abundance relative to hydrogen in the equatorial regions is about the same as that in polar caps (see figure 7) and the hydrogen emission lines intensities are similar in both regions. This indicates that the polar O III features are caused by different ionization conditions in the polar and equatorial regions. To reproduce this scenario, with constant photoionization conditions models, the second component (globules) was constrained in a solid angle of 90 degrees for each polar cap.

The structured shell models have a condensed mass fractions $fc$ between 0.1 and 0.7 and a total mass between $1 \times 10^4$ and $3 \times 10^3$ solar masses. Fixed shell geometry was assumed and the average globule density contrast relative to the background density varied, as well as the total shell mass to produce a grid of model shells. There are other free parameters, namely the central source temperature and luminosity. The density in the clumps was constrained to the range of 3 to 50 times the background density and their size between $10^{15}$ cm to $10^{16}$ cm, as suggested by the emission line maps.

A central source effective temperature of 63000 K and luminosity $\log(L) = 35.5$ erg.s$^{-1}$ were found in the best fit clumpy model. The total shell mass for this model is $9 \times 10^4$ M$_{sun}$. The mean electron temperature derived is 5900 K and the electron density varies from a maximum of 6000 cm$^3$ in the clumps to 400 cm$^3$ in the background regions. The best fit shell has a condensed mass fraction of 0.56. The presence of globules in the shell significantly improves the lower ionization line ratios. The Nitrogen forbidden lines, such as [N II] $\lambda\lambda$6548,6584, have a consistent behavior, with intensities comparable to H$\beta$ which is much better than the homogeneous model. Even for the lower $fc$ values reasonable [N II] intensities were obtained. Denser clumps (20 to 50 times the background) with a higher condensed mass fraction, $fc$, result in a smaller [N II] intensity. Less dense clumps (3 to 15 times the background) produce higher [N II]/H$\beta$ ratios. This occurs because in higher density clumps the N II is de-excited by collisions. The [N II] line intensity also depends on the total shell mass for a fixed geometry since a high mass shell is also a denser shell.

The clumps also improve the modeling of the H$\beta$ emission. An increase in the H$\beta$ flux by a factor 3 is found when compared to the homogeneous shells with equal central source parameters. This means that the shell mass could be around half of that predicted by the homogeneous models. The [O III] intensities were also improved by the presence of clumps in the shell. The clumpy model also yields a better [O II]/H$\beta$ ratio, within 60 % to 100 % of the observed value. The main parameters affecting this ratio are the shell mass and $fc$. For a shell where $fc$ = 0.5 the [O II]/H$\beta$ ratio is 10 times larger than the ratio of homogeneous models. A clumpy shell with mean density above 1000 cm$^3$ is required to have a good fit of the [O II] lines. The derived [O I]/H$\beta$ ratios in our clumpy models are much better than those found in the homogeneous model, with a match within 10 % of the observed ratios. The He I $\lambda$5876/H$\beta$ line ratio has an excellent fit in the clumpy model. This particular line ratio does not present large variations with the central source temperature (10% from 50000 to 100000K). It depends mostly on the shell density and therefore on $fc$. Large $fc$ values tend to increase this line ratio.

In all clumpy models a common problem is faced when fitting the He II intensities, which could not be reproduced simultaneously with the [O III] line fluxes. Trying to fit the He II/H$\beta$ ratio alone, would require a central source of around 92000K. However, for such a hot central source the O III lines would have 10 times the observed intensities, and this would be inconsistent with the Oxygen abundance determinations. We decided to look at the He II 4686/ He I 5876 ratio behavior in the grid of models to find which physical parameters are relevant. As expected, such a line ratio has a strong dependence on the temperature and luminosity of the ionizing source and also on the shell density (or $fc$). Higher temperatures and luminosities mean large line ratios, since we have an increase in the He$^+$ ionization. An increase on $fc$ and shell density diminishes the He II/ He I ratio because of a less effective ionization inside the high density clumps. For instance, if $fc$ = 0.3, a ratio He II$\lambda$4686/H$\beta$ is obtained, which is 50% below the result from a homogeneous model with the same parameters. The Oxygen abundance in the shell also affects the He II/He I ratio. Higher Oxygen abundance decreases this line ratio. A family of solutions with lower Oxygen abundances (log(O/H) = -4.3) exists, although this value is far from what is found in section V. As expected, higher Oxygen abundances also produce cooler shells.

V.3.3 Disk-like central source

The aspherical ionization of a novae remnant was first suggested by Petitjean et al. (1990) and Gill & O'Brien (2000) to explain some features of DQ Her and FH Ser shells. In the present work the irradiation of the shell by a disk shaped ionizing source was simulated. If at least part of the ionizing radiation field comes from the inner part of the accretion disk, the polar regions are expected to receive a larger ionizing flux than the equatorial regions of the shell. This is due to the source aspect variation from polar to equatorial regions and self-absorption by an optically thick accretion disk.

The models with a disk shaped central source are computed considering the total luminosity as in the spherical central source clumpy models (grid II models hereafter) described in sections V.3.2 and V.3.1 while the flux decreases with the aspect of the inner disk as "seen" from the shell at given angle relative to the polar axis. In the equatorial regions an edge on disk is "observed". Therefore we could compound a model weighted by the lines luminosity contribution of many sub-models. The sub-models used have the same shell mass, chemical abundances and temperatures, but they have different geometries, condensed mass fraction and central source luminosities. The density profile was defined by the 3D shell geometry of a closed hourglass with spherical globes. Therefore at each direction the shell extends out to a different distance from the central source. The sub-models were weighted using the area defined by the solid angle of each direction. These models are named grid III. The ionizing radiation field was assumed to be a high gravity hot star atmosphere. Ionizing spectra from Rauch (2003) hot stellar atmospheres with log(g) = 7-8 and temperatures from 50000K to 150000K were used in the calculations.

The spatial resolved spectra showed that the shell can be divided in two main regions. The first one is the polar caps, where we have most of the [O III] and [N II] emissions. The He II lines also have their major emission coming from the polar regions. However, the second one, namely the equatorial ring region, shows a small contribution to these lines. The [O III] line emission for example, is five times less dim than in the polar regions. There are other features indicating that the equatorial regions have a less intense ionizing flux, including the abundance independent O II/O III and He I/ He II ratios that are larger in the equatorial regions.

Our calculations showed that grid III models yield a much better description of both [O III] and He II lines. The best fit model has a central source temperature $T_{eff}$ = 65000K and a total luminosity log(L) = 36 erg.s$^1$, which are in agreement with the UV observations (Selvelli and Friedjung, 2003). In this model the [O III]/Hβ ratio is 1.5 and He II 4686/Hβ is 1.2. Both are close to line ratios measured in the HR Del shell. The polar regions have a photon flux nearly 20 times greater than the equatorial regions. Because of this feature, most of the He II λ4686 flux comes from the polar regions, as observed on the He II line maps (figure 12). The [O III] line intensities in polar regions are 5 times greater than the equatorial regions, which is again in agreement with the shell measurements. The [O II] λ7325 blend and the [O I] λ6300 intensities obtained in grid III models are the same as in grid II. This means that the presence of the clumps is still important to reproduce the lower ionization lines. The best fit grid III model has a lower density background, with mean value of 400 cm$^3$ and a clumpy component with mean value of 2000 cm$^3$. The density of the shell has a range from 200 cm$^3$ to 6000 cm$^3$. The lower density regions contribute to most of the He II emissions and the higher density globules contribute to the [O III] and [N II] emission. The total Hβ flux displays small differences when compared to grid II. Therefore the total shell mass remained roughly the same as in grid II models. The condensed mass fraction of the best fit model in grid III is $fc$ = 0.7 being clumpier than the best fit models in grid II ($fc$ = 0.56).

## VI – Discussion

### VI.1 Shell morphology, ionization and chemical abundances

The 2D spectroscopy of the shell (see figures 2 and 3) shows many structures, like polar caps, equatorial rings, clumps all over the shell, and also reveals the shape of the outer part of the shell. The image of HR Del shell in the light of Hα + [N II] lines shows a different outer aspect ratio (q = 1.6) when compared to the shell in the [O III] lines (q = 1.95). This feature was also observed by Harman & O'Brien (2003) who analyzed the HST images in [N II], Hα and [O III] filters. Line dependent morphology was observed in other novae, like DQ Her (Baade 1940) and it seems to be related to the speed class of the nova. Fast novae, with a massive white dwarf, have a spherically shaped ejecta, while slow novae in which lower mass white dwarfs are expected, show a prolate ejecta. However, there are exceptions like DQ Her, which has a mass white dwarf lower ($M_{WD}$ = 0.6 $M_{sun}$) than HR Del, but presents an axial ratio of 1.4 (Slavin et al. 1995). The Balmer lines are less affected by the aspherical illumination and their intensity maps are consistent with a density distribution without polar enhancements. On the other hand, our Oxygen and Nitrogen abundance analysis shows that the average polar and equatorial abundances are similar. The He abundances in those regions are comparable as well. Therefore, the high [O III] and [N II] emissions observed in the polar caps suggest that the ionizing flux in the equatorial regions is less intense than in the polar regions. The He II emission decay with distance is faster in the equatorial regions than it is in the polar regions (see figure 6). All the line ratios observed in both regions point towards an aspherical illumination, consistent with the scenario of a disk shaped central source. This effect is not directly related to the speed class of the nova.

The 2D spectroscopy of the shell (see figures 2 and 3) show many structures, like polar caps, equatorial rings, clumps all over the shell, and also reveals the shape of the outer part of the shell. The line maps close to the rest wavelength of H$\alpha$ in figure 3 (frames FGH) show that there are two ring structures and that they have symmetrical velocities relative to the central object. Smaller and less intense rings could be seen at [N II] wavelengths (frames C and L). The redshifted and the blueshifted structures have equal dimensions and they correspond to the outer part of the shell. The projection of these structures on the plane of the sky is a circular ring, with a diameter of 8" and thickness of 1.7". This means that these are spherically shaped structures, located at each side of the central object and look like a closed hourglass with the velocity of each structure being proportional to the distance of the center. The closed hourglass shell model, first proposed by Harman and O'Brien (2003), describes the outer part of the observed shell in HR Del. The shaping of nova remnants by binary motion and rotating white dwarf envelopes was modeled by Lloyd et al. (1997) and Porter et al. (1998). Their 2.5D hydrodynamic simulations showed that in slow novae, which also have slow expansion velocities, the binary motion produces polar caps. The angular momentum of the envelope, which depends on the rotational velocity, is responsible for producing a prolate shell. The HR Del shell shows both features, not only a prolate shell but also blobs in the polar regions. However, their models do not reproduce exactly the external structure of the HR Del shell, like the equatorial enhancements and the polar caps. The equatorial density enhancements seem to be an outer part of the shell superimposed in the line of sight. It was not also observed the tropical rings seen in the synthetic images derived from the hydrodynamic solutions for slow novae. Maybe a full 3D hydrodynamic model could better reproduce the observed shell. Gill and O'Brien (1999) used the Porter et al. (1998) shell models to obtain their line profiles for a shell observed at different inclinations. Our data show the H$\beta$ line profile with 4 peaks of about the same intensity which is not reproduced by their line profile models. The H$\beta$ line profile of our data is similar to the [O III] $\lambda$4363 profile observed in HR Del spectra in October of 1969 by Rafanelli and Rosino (1978). The [O III] line profile itself has changed over the years. In 1969 the four [O III] peaks had the same intensity. In the 1978-1980 observations by Küster and Barwig (1988) the intensity of the higher velocity components were larger and further observations by Solf (1983) confirm that the intensities of the lower velocity components were decreasing. Our data do not show any lower velocity component contribution to the [O III] line profiles. We only see the high velocity emission component that comes from the polar regions. This could mean that the ionization flux decreased over the time at the lower velocity (equatorial) regions as the white dwarf was cooling and the accretion disk became the most important ionization source.

The polar caps structures are inside the hourglass observed in the Balmer lines and closer to the ionizing source. Our velocity measurements suggest that the hourglass structures have a slightly higher expansion velocity when compared to the polar caps. The HR Del light curve shows that the nova had two main optical flux peaks, one at the maximum visible light in December of 1967 and other in May of 1968 (Rafanelli and Rosino, 1978). This could mean that there were two different mass ejections events and could explain why we see these two different structures at distinct distances from the WD. Early observations of RS Oph 2006 outburst (day 13 to 155 after eruption) showed that this recurrent nova ejecta also has a bipolar structure (O'Brien et al. 2006 and Bode et al. 2007). The models by Bode et al. (2007) for the [O III] emission in RS Oph (their figure 2) is surprisingly similar to what we see on our H$\alpha$ line maps. These authors argued that the wind of the evolved secondary is denser at the equatorial regions decelerating the material ejected from equatorial regions, given the ejecta a bipolar shape. However, it is not clear how this interaction could generate the spherically shaped components of the shell.

Porter et al. (1998) suggested that the angular momentum of the accreted material leads to a shear-induced mixing at the orbital plane region of the white dwarf. If this process is more important than the particle diffusion then one should observe abundance gradients in novae remnants. Our spatially resolved spectral data do not show abundance gradients between polar and equatorial regions as expected if the shear-mixing occurs. Fast WD rotation at a significant fraction of the surface Keplerian value could inhibit the mixing. However, abundance gradients were measured in radial directions. The outer part of the shell seems to be less enhanced in Oxygen and Nitrogen than the inner part of the ejecta. This feature can be explained if the mixing is more effective at the bottom of the accreted envelope when compared to its surface.

VI.2 Photoionization models

Homogeneous photoionization models with a spherically symmetrical ionization source could not reproduce the spectra of the HR Del shell. The major problem was the fit of lower ionization stage lines, such [O II], [O I] and also the [N II] line intensities. These lines arise from denser and cooler regions, that only could exist within a clumpy shell. As expected, the clumpy shell model yields a much better fit of these lines. The presence of condensations also improves the match of the H$\beta$ flux for a model with equal shell mass and central source parameters. A side effect of increasing the condensed mass fraction is a decrease in the He II/H$\beta$ ratio. For $fc$ = 0.3 this ratio was decreased by 50% when compared to a homogeneous shell. The He II lines arise from low density regions and/or regions that are close to the central source. This is illustrated in figure 12 which shows that the He II emission comes from regions not correlated with the polar caps. It is also possible to see that the He II diffuse emission is elongated and has an axial ratio similar to the [O III] line map shell. The analysis of the emission line kinematics suggests that there are many more small clumps than large ones, but this analysis is limited by the spatial resolution of the line maps. Lloyd et al. (1997) obtained that the clump size should be related to the speed class of the nova. They showed that for slow novae the maximum size of the clumps is much smaller than those in fast novae shells. The clumps are believed to be formed by the Rayleigh-Taylor instability (Shara 1982) and for a slow novae ejecta the maximum size of the instabilities is roughly 1% of the shell radius. Experimenting with several clumps size distributions modest variations were obtained in the line ratios of about 10%. The intensity of the O III $\lambda$ 5592 observed in the inner parts of the shell may suggest that there are neutral clumps in this region since this transition is boosted by charge transfer from H I at neutral fractions above $10^{-4}$ (Clegg and Walsh, 1985).

The spherically symmetrical ionization source models with condensations fail in reproducing simultaneously the He II and [O III] emission lines. The presence of clumps in the shell slightly improves the He II/[O III] because the He II luminosity is more affected by the ionization source luminosity which can be larger in clumpy models. The clumps prevent the O$^{+2}$ ions to be ionized and the [O III] luminosity has a small variation with increasing ionization photons flux. The fitting of both He II $\lambda$4686 and [O III] $\lambda$5007 line problem was also observed in other novae models, such V1974 Cyg by Vanlandingham et al. (2005). These authors also used the code CLOUDY models and their best fit could not reproduce both He II and [O III] lines.

The model with a disk shape ionizing source produced an illumination variation from polar to equatorial regions. These models resulted in the best fit for both He II and [O III] simultaneously. Most of the He II flux comes from the polar axis directions, where the ionizing flux at high energies was larger. Besides, most of the [O III] luminosity also comes from the polar regions. This line is less affected by the central luminosity than He II λ4686. The model fluxes at the shell equator are zero in the aspect approximation for a geometrically thin disk. However, a well-developed accretion disk gradually increases the UV opacity towards the equatorial direction. The models used in this work do not account for the disk limb darking, which may be significant in the UV (Diaz et al. 1995). The disk flare and its resulting shadow close to the shell equator were also neglected. This approximation is also linked to the lack of non-radial transfer of diffuse radiation in the shell.

Although the photoionization models explain most of the spectral features, there are other physical effects that can contribute to the observed spectrum. Contini and Prialnik (1997) showed that the spectral features of the recurrent nova T Pyx could be explained by the contribution of shocks as a mechanism of line excitation. These authors showed that shocks can excite the neutral and less ionized ions, such [O I], [O II] and [S II] and enhanced their emission lines. The HR Del ejecta shows several structures at different velocities. The post-maximum light curve indicates that more than one mass ejection event could have happened during the nova outburst. P-Cygni profiles in the UV lines suggest high velocity winds (Selvelli and Friedjung, 2003) and accretion disk wind models require that the mass loss in the wind corresponds to 2-10% of the accretion rate (Puebla et al, in preparation). On the basis of these considerations, shocks may play an important role in the energy budget of the shell. In fact, the HR Del spectrum show enhanced intensities in some low ionization lines, such as [O I]λλ 6300,6364 and the [S II]λλ 6716, 6731 doublet, that could not be explained by the photoionization model alone. A detailed modeling of the shocks in this system may be pursued using higher S/N 2D spectroscopy of lines that are potentially excited by shocks combined with precise velocity measurements of the outer shell.

VII – Conclusions

The very slow novae remnant of HR Del was observed with IFU-GMOS at Gemini north revealing the emissivity and velocity structure of the shell in several emission lines. The spectral data were used to revise the distance to this object by expansion parallax as 850(60) pc. The expansion velocities of 560 (60) km.s$^1$ for the polar regions and 300(60) km.s$^1$ for the equatorial region using the [O III] line profiles were also obtained. The outer shell expands at 630(60) km.s$^1$ as derived from the Hα line. The prolate shell has an axial ratio of 1.6 in the Balmer lines and 1.9 in the [O III] lines, following the trend of slow novae shells. The ejecta were found to be very clumpy on all line maps. Some clumps seem to be located outside the shell. These runaway clumps have the low ionization emission enhanced, consistent with a higher density environment when compared to their neighborhood environment. Another conclusion is that there are more small clumps than large ones, but this analysis is still limited by the spatial resolution of the line maps. The spatial spectral analysis shows that the polar regions are more ionized than the equatorial ones, which is consistent with the hypothesis of aspherical illumination by the inner accretion disk.

The kinematic analysis showed that the outer part of the shell is defined by a closed hourglass shape. The Balmer lines define the outer part of the shell, but a smaller and faint ring structure in the [N II] emission lines was also observed. There are polar caps located inside the hourglass structure. Each polar cap is within a 90 degree cone, centered in the ionizing source. These polar caps were seen only in the [O III] and [N II] lines, that are the high ionized ions strongest lines. The equatorial emission enhancement seen on the Balmer line maps is caused by the ring structures that are superimposed in the line of sight because of projection effects. The presence of both outer rings and polar caps suggests that there were two mass ejection events, which is also supported by the light curve of HR Del. The shaping of binary remnants by binary motion and rotating WD envelopes as predicted by 2.5D hydrodynamic models could not reproduce the observed HR Del shell features and also the emission line profiles.

Abundance gradients between polar and equatorial regions could not be found as expected if significant shear-mixing by the accretion disk had occurred before the eruption. On the other hand, the analysis in this study suggests that there are Oxygen and Nitrogen abundance variations between inner and outer regions of the shell. This is consistent with a more effective mixing at the bottom of the accreted envelope.

All the observed ejecta features were used to build a 3D model of the shell, with clumps and a closed hourglass mass distribution. Unlike a spherically symmetrical shell, our 3D model reproduces the lower ionization state lines such [O I] and [O II] and also the [N II] and [OIII] lines. Average shell electron temperatures from [O I] lines are around 6500(500) K, in agreement with those found using the [O III]$\lambda\lambda$4959,5007/4363 ratio. The temperature inside the clumps can be as low as 5700(500) K. Both symmetric and 3D clumpy models could not fit simultaneously the He II and [O III] lines using the abundances obtained from literature and our data. A family of 3D clumpy models including a disk shaped central source disk improved significantly the fit of both He II and [O III] lines. Clumpy shell models also yields lower shell mass estimates. The luminosity obtained for the central source is $10^{36}$ erg.s$^{-1}$ with an effective temperature of 65000K. A high shell mass (1.24 x 10$^{-3}$ solar masses) was estimated using the integrated H$\beta$ luminosity, which is larger than accreted mass required to trig the TNR at the low mass white dwarf (~1 x 10$^{-4}$ M$_{sun}$) obtained by Yaron et al. (2005) grid of models. The best fit 3D models indicate the presence of condensations with a density contrast up to a factor of 30 relative to the lowest density regions. A shell where 50% to 70% of the mass is contained in clumps is proposed by the simulations of HR Del ejecta. Some spectral features could not be exactly reproduced by the photoionization models alone. The ejecta velocity field, the outburst and the fast wind inferred by UV spectrum suggest that shock may occur and contribute to the observed spectrum features, like [O I] and [S II] enhanced intensities.

ACKNOWLEDGMENTS

We acknowledge the support from CNPq under fellowship #142389/2006-4 and grant #305275. We also thank the referee for providing valuable comments and help improving the manuscript.

This work was based on observations obtained at the Gemini Observatory, (GN-2002A-Q11), which is operated by the Association of Universities for Research in Astronomy, Inc., under a cooperative agreement with the NSF on behalf of the Gemini partnership: the National Science Foundation (United States), the Science and Technology Facilities Council (United Kingdom), the National Research Council (Canada), CONICYT (Chile), the Australian Research Council (Australia), Ministério da Ciência e Tecnologia (Brazil) and Ministerio de Ciencia, Tecnología e Innovación Productiva (Argentina)

*References:*

| | |
|---|---|
| *1994ApJ...432..427* | *Aller, Lawrence H.* |
| *1989MNRAS.238...57* | *Almog, Yael; Netzer, Hagai* |
| *1940PASP...52..386* | *Baade, W.* |
| *1990ApJ...354..529* | *Bertoldi, Frank; McKee, Christopher F.* |
| *2004ASPC..313..504* | *Bode, M. F.* |
| *2007ApJ...665L..63* | *Bode, M. F.; Harman, D. J.; O'Brien, T. J.; Bond, Howard E.; Starrfield, S.; Darnley, M. J.; Evans, A.; Eyres, S. P. S.* |

*1982PASP...94..916*            Bruch, A.
*1987MNRAS.229P..31*            Clegg, R. E. S.
*1987MNRAS.228..759*            Clegg, R. E. S.; Middlemass, D.
*1985MNRAS.215..323*            Clegg, R. E. S.; Walsh, J. R.
*1997ApJ...475..803*            Contini, M. ; Prialnik, D.
*1987JRASC..81..195*            De Robertis, M. M.; Dufour, R. J.; Hunt, R. W.
*1995ApJ...452..704*            della Valle, Massimo; Livio, Mario
*1995AAS...186.0911*            Diaz, M. P.; Wade, R. A.; Hubeny, I.
*2001ASPC..247..227*            Diaz, Marcos
*2000AJ....120.2007*            Downes, Ronald A.; Duerbeck, Hilmar W.
*2003MNRAS.340.1136*            Ercolano, B.; Barlow, M. J.; Storey, P. J.; Liu, X.-W.
*1986ApJ...310L..67*            Ferland, G. J.
*2005AAS...206.3106*            Ferland, G. J.
*1982A&A...114..351*            Friedjung, M.; Puget, P.; Andrillat, Y.
*1993prpl.conf...75*            Gehrz, Robert D.; Truran, James W.; Williams, Robert E.
*2000MNRAS.314..175*            C. D.; O'Brien, T. J.
*1999MNRAS.307..677*            Gill, C. D.; O'Brien, T. J.
*2003MNRAS.344.1219*            Harman, D. J.; O'Brien, T. J.
*1980PASP...92..458*            Hutchings, J. B.
*1988A&A...199..201*            Kuerster, M.; Barwig, H.
*2003A&A...405..189*            Lim, A. J.; Mellema, G.
*1997MNRAS.284..137*            Lloyd, H. M.; O'Brien, T. J.; Bode, M. F.
*1998A&A...331..335*            Mellema, G.; Raga, A. C.; Canto, J.; Lundqvist, P.; Balick, B.; Steffen, W.; Noriega-Crespo, A.
*2005AIPC..804...44*            Morisset, C.; Stasi ska, G.; Peña, M
*2006Natur.442..279*            O'Brien, T. J.; Bode, M. F.; Porcas, R. W.; Muxlow, T. W. B.; Eyres, S. P. S.; Beswick, R. J.; Garrington, S. T.; Davis, R. J.; Evans, A.
*Astrophysics of gaseous nebulae and active galactic nuclei (2ND Edition) / University Science Books, 2006*     Osterbrock, Donald E.; Ferland, Gary J.
*1990A&A...240..433*            Petitjean, P.; Boisson, C.; Pequignot, D.
*1998MNRAS.296..943*            Porter, John M.; O'Brien, T. J.; Bode, Mike F.
*1978A&AS...31..337*            Rafanelli, P.; Rosino, L.
*2003A&A...403..709*            Rauch, T.
*2003A&A...404..301*            Ritter, H.; Kolb, U.
*1972BAAS....4..217*            Sanyal, Ashit
*2003A&A...401..297*            Selvelli, P.; Friedjung, M.
*1981ApJ...243..926*            Shara, M. M.
*1982ApJ...261..649*            Shara, M. M.
*1982ApJ...258L..41*            Shara, M. M.; Moffat, A. F. J.
*1994ASPC...61..327*            Shaw, R. A.; Dufour, R. J.
*1993AJ....106.2408*            Shore, Steven N.; Sonneborn, George; Starrfield, Sumner; Riestra-Gonzalez, Rosario; Ake, T. B.
*1994MNRAS.266L..55*            Slavin, A. J.; O'Brien, T. J.; Dunlop, J. S.
*1995MNRAS.276..353*            Slavin, A. J.; O'Brien, T. J.; Dunlop, J. S.
*1983ApJ...273..647*            Solf, J.
*1978AcA....28..333*            Tylenda, R.
*2005ApJ...624..914*            Vanlandingham, K. M.; Schwarz, G. J.; Shore, S. N.; Starrfield, S.; Wagner, R. M.
*1994ApJS...90..297*            Williams, R. E.; Phillips, M. M.; Hamuy, M.
*1992AJ....104..725*            Williams, Robert E.
*1994ApJ...426..279*            Williams, Robert E.
*2005ApJ...623..398*            Yaron, O. ; Prialnik, D. ; Shara, M. ; Kovetz, A.

FIGURES

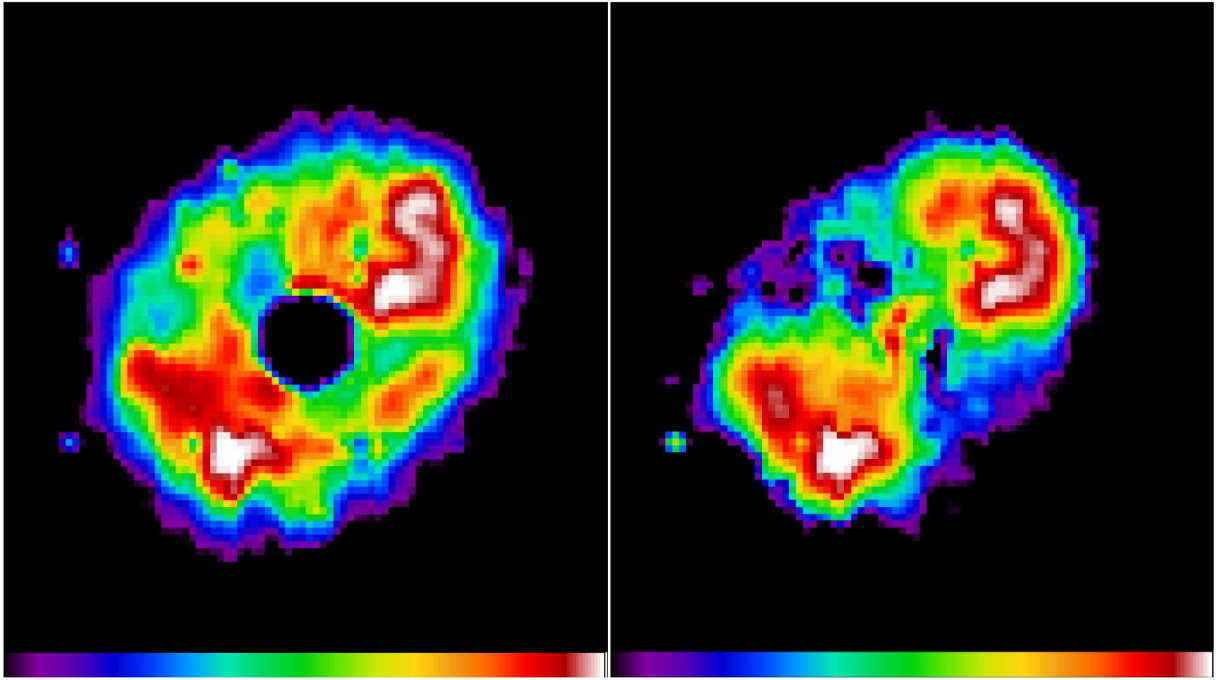

*Figure 1 - Maps of HR Del remnant shell in Hα + [N II] (left) and [O III]λ5007 (right). The map of Hα + [N II] had the central emission contribution area removed to improve the shell contrast. Both images have equal intensity scale from 1 x 10$^{-15}$ to 4 x 10$^{-14}$ in erg.s$^{-1}$.cm$^{-2}$ and had the continuum subtracted.*

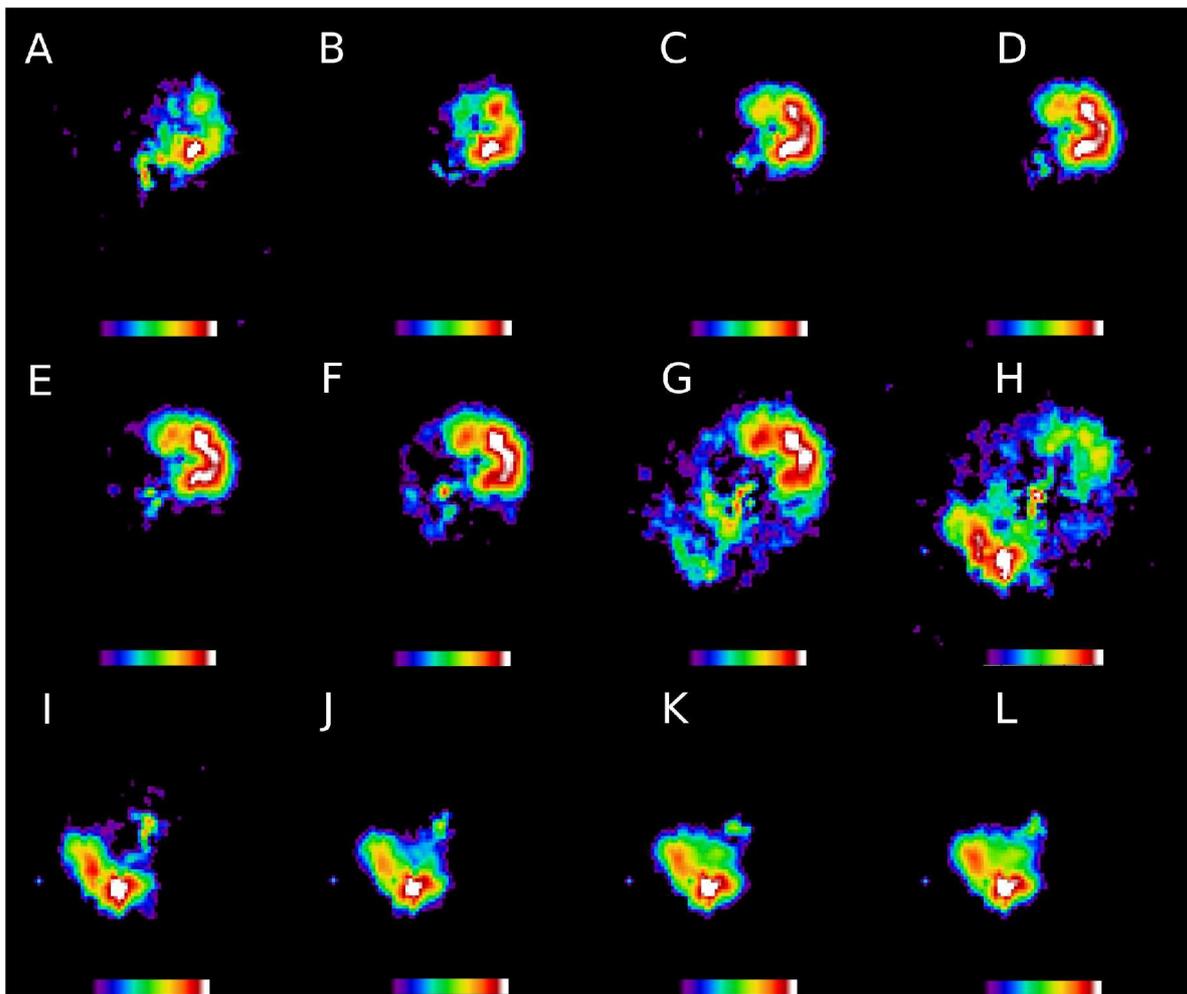

*Figure 2 – Maps of [O III] λ5007 from -900 km.s$^{-1}$ to 540 km.s$^{-1}$ from panels A to L with steps of 120 km.s$^{-1}$ or 2 Angstrons. At the top there is the blueshifted polar cap and at the bottom the red shifted one. All these maps have equal scale intensity from 1 x 10$^{-16}$ to 2 x 10$^{-15}$ erg.s$^{-1}$.cm$^{-2}$ and same velocity range.*

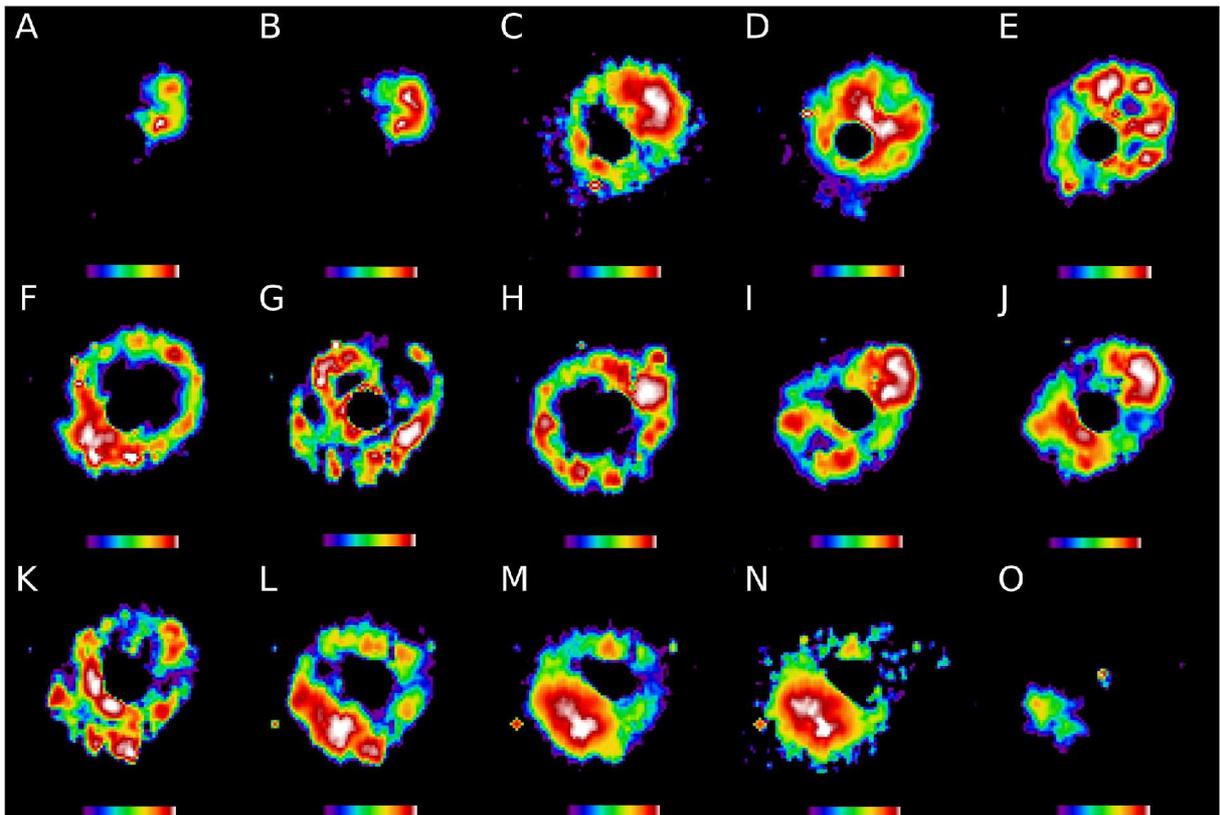

*Figure 3 - Maps of Hα + [N II]λλ 6548,6584 from -1510 to 1940 km.s$^{-1}$ from panels A to O. Each step corresponds to 230 km.s$^{-1}$ or 5 Angstrons. All these maps have the same intensity scale from 10$^{-16}$ to 5.3 x 10$^{-15}$ erg.s$^{-1}$.cm$^{-2}$ and same velocity range. The central source and surrounding regions were removed from all maps to increase the image contrast. All images were continuum subtracted.*

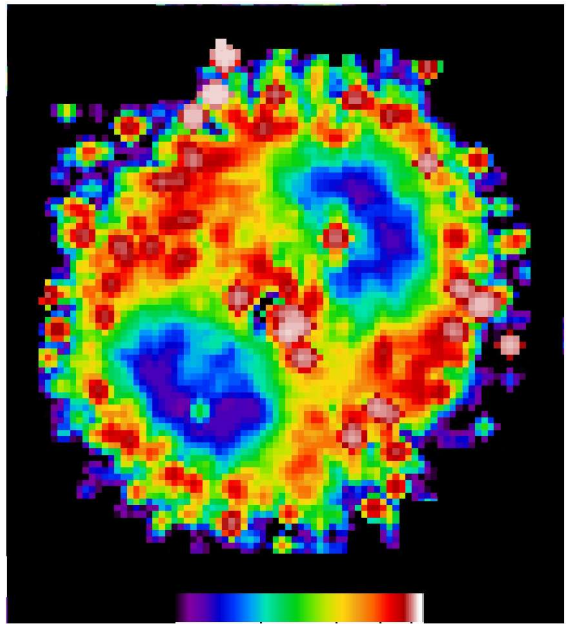

*Figure 4 – The Hα+[N II]/[O III]λ5007 ratio emphasize the differences between this emission lines. Both lines had the continuum subtracted. A reddening correction of E$^{(B-V)}$ = 0.16 was applied. The color scale is from 0.5 to 40.*

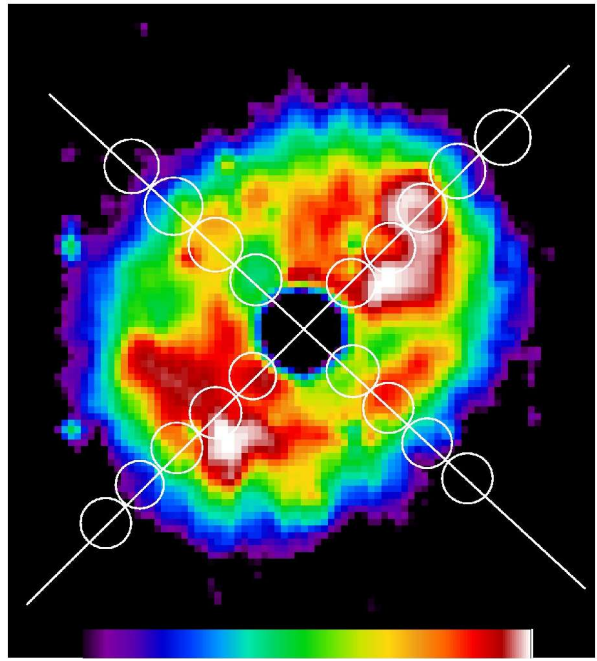

*Figure 5 –  Broad Hα + [N II] image of HR Del shell with diagnostic regions marked along the polar and equatorial axis. Color scale is from $10^{-15}$ to $4 \times 10^{-14}$ erg.s$^{-1}$.cm$^{-2}$.*

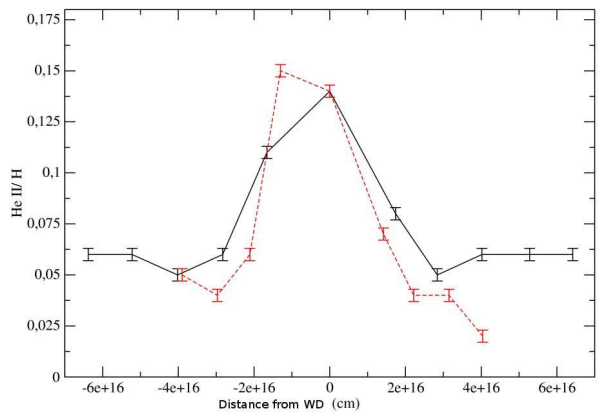

*Figure 6 – He+ ion abundance in polar and equatorial axes. For both directions and axis the decay is close to $r^2$. The positive value of axis corresponds to the blue shifted side in the polar axis and the bottom part of the shell in the equatorial axis.*

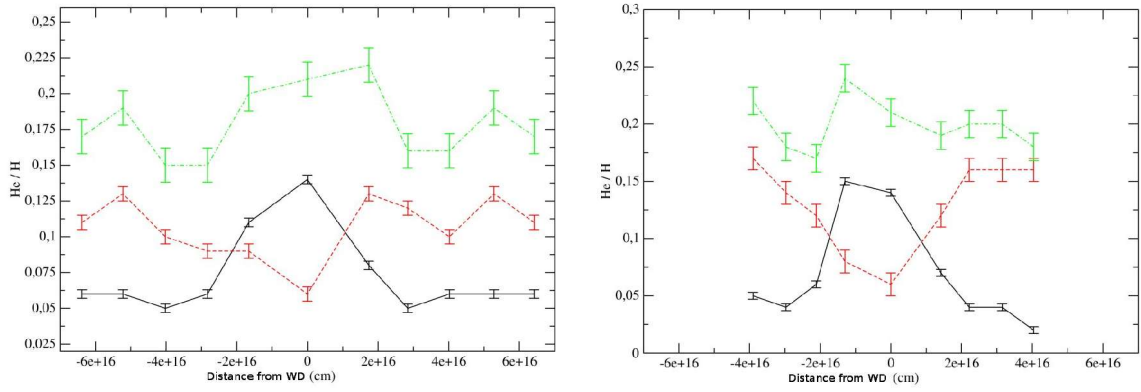

*Figure 7 – Helium ion and total abundance profile in the polar and equatorial axes. The He II λ4686 and He I λ5876 lines were used to determine the ions abundance of $He^{++}$ and $He^{+}$, respectively.*

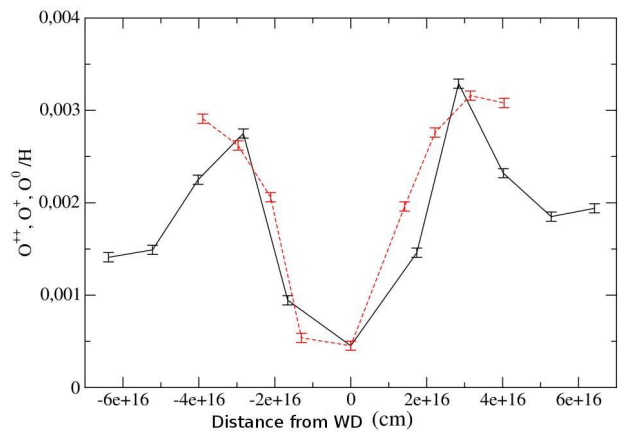

*Figure 8 – Oxygen ions abundances in the polar and equatorial axis. The abundances were obtained using [O III] λ5007, [O II] λ7325 blend and [O I ] λ6300 line fluxes. It was not possible to measure the $O^{+3}$ abundances.*

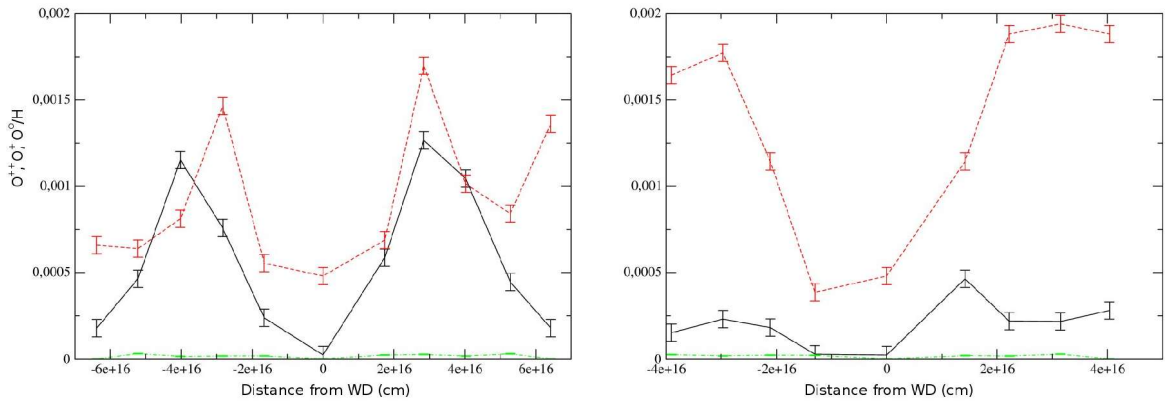

*Figure 9 – Oxygen ion abundances along the polar (left) and equatorial axes (right). The continuous line is $O^{+2}$, the dotted line $O^{+1}$ while the dotted is $O^{0}$. It was not possible to measure the $O^{+3}$ ions.*

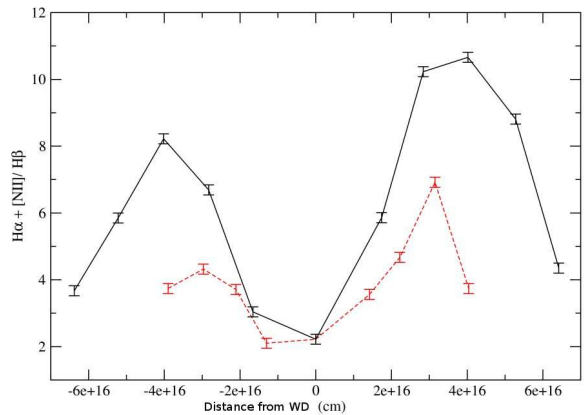

*Figure 10 – Hα + [N II] / Hβ ratio along the polar and equatorial axis. The continuous line represents the polar axis direction, and the dotted, the equatorial axes. The largest line ratios occur at the clumpy regions where the NII population is greater than in the regions next to the central source.*

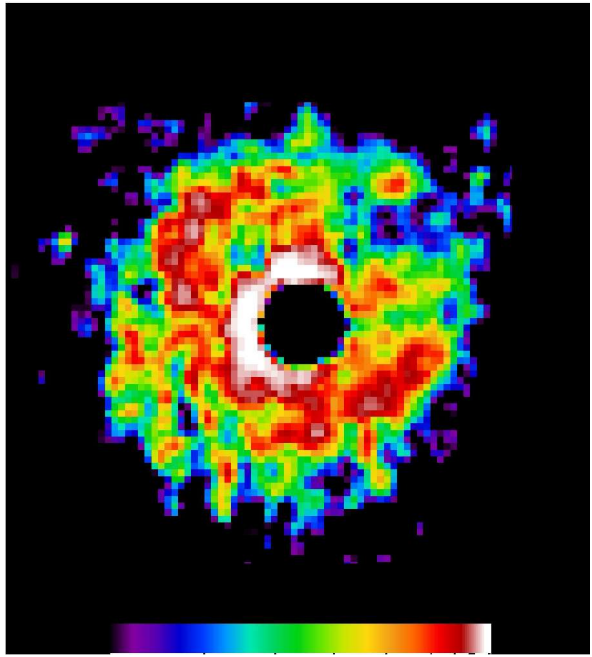

*Figure 11 – Hβ line maps with central wavelength at 4861 A and a width of 10 A. Most of the emission comes from regions where the ring structures are overlapping at the equatorial axis. The color scale is from $2 \times 10^{-16}$ to $5 \times 10^{-15}$ erg.s$^{-1}$.cm$^{-2}$.*

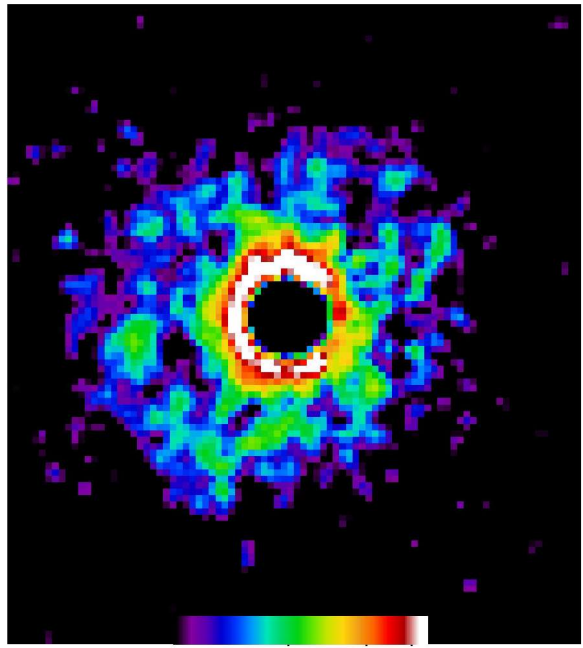

*Figure 12 – He II λ4686 emission map, with a width of 10 A. It is possible to observe that the shell has a large axial ratio, with an extended polar emission region. However, the He II emission regions are not correlated with the polar caps seen on [O III] line maps. The most of the emission comes from the inner regions of the shell. The color scale is form $10^{-15}$ to $4 \times 10^{-14}$ erg.s$^{-1}$.cm$^{-2}$.*

TABLES:

| line ID | total | central | shell |
|---|---|---|---|
| HI 4341 | 36.97 | 34.54 | 2.52 |
| CIII+NIII 4649 | 75.23 | 67.65 | 7.72 |
| HeII 4686 | 136.97 | 90.76 | 46.39 |
| [FeIII](3F) 4702 | | | 3.77 |
| [NeIV](IF) 4714-20 | | | 4.32 |
| [FeII](4F) 4728 | | | 1.67 |
| [FeIII](3F) 4734 | | | 1.10 |
| [FeIII](3F) 4755 | | | 2.87 |
| NII(20) 4780 | | | 0.87 |
| [FeII](4F) 4798 | 3.44 | 0.68 | 2.66 |
| NII(20) 4803-10 | 3.75 | | 0.85 |
| [FeIII](3F) 4821 | 1.59 | 0.50 | 1.09 |
| FeII(30) 4838 | 4.83 | | 2.87 |
| HI 4861 | 100.00 | 53.95 | 45.88 |
| [FeVII](2F) 4894 | 6.45 | | 1.98 |
| [FeII](20F) 4905 | 7.83 | | 3.45 |
| HeI 4922 | 8.91 | 5.27 | 3.67 |
| [FeIII](1F) b 4936 B | 0.82 | | 0.24 |
| [FeVII](2F) 4941 B | 2.94 | | 1.48 |
| [OIII](1F) 4959 | 36.58 | 8.10 | 28.74 |
| [OIII](1F) 5007 | 113.45 | 29.20 | 84.87 |
| I 5029? NIII | 3.68 | 0.76 | 2.82 |
| [FeII](4F) 5036: | 0.74 | | 0.60 |
| SiII(5) 5041 | 2.21 | | 1.51 |
| HeI(47) 5047 | 1.92 | | 1.01 |
| FeII(49) 5235, FeVI(1F)5237 | 5.36 | 3.52 | 1.86 |
| [FeIII](1F) 5270 | 1.23 | 0.57 | 0.63 |
| HeII 5411 | 10.42 | 6.46 | 4.08 |
| [OI](3F) 5577, [FeII](2F)5582 | 4.82 | 1.46 | 3.36 |
| OIII(5) 5592 | | 1.68 | |
| [CaVII](1F) 5620 | 5.03 | 0.74 | 3.92 |
| [CaVII](2F) 5631, [FeVI](1F) 563? | 2.85 | 0.52 | 2.28 |
| NII(3) 5676 | 1.53 | 1.02 | 0.51 |
| [FeII](33F) 5684, NII(3) 5686 | 1.96 | | 1.44 |
| CIV 5802-12 | 13.98 | 13.53 | 0.46 |
| HeI 5876 | 12.93 | 4.55 | 8.49 |
| [OI](1F) 6300 | 2.72 | 0.59 | 1.40 |
| SiII(2) 6347 | 1.06 | 0.44 | 0.40 |
| [OI](1F) 6364 | 1.32 | 0.18 | 1.15 |
| SiII(2) 6372 | 0.39 | 0.21 | 0.15 |
| FeII(74) 6457 | 0.37 | 2.18 | |
| HI 6563+[NII] 6565 | 387.39 | 154.75 | 232.65 |
| HeI 6678 | 10.24 | 9.50 | 0.75 |
| HeI(10) 7065 | 2.61 | 2.39 | 0.46 |
| HeI 7281 | 1.69 | 1.67 | 0.06 |
| [OII](2F) 7320+7331 | 3.13 | 1.47 | 1.49 |
| CIV 7705-08-26 | 8.92 | 4.50 | 2.80 |
| MgII(8) 7877 | 0.80 | 0.35 | 0.50 |

*Table 1 – Line identification and line intensities relative to Hβ. The column named Total means the flux of all object, ionization source and nebulae, the central column is the ionization source regions and the shell column is the nebular emission only.*

| Lines | Lines |
|---|---|
| C IV 1549 | C III 5826 |
| He II 1640 | He I 5876 |
| H 1 4340 | [Fe VII] 6087 |
| [OIII] 4363 | [O I] 6300 |
| N III 4641 | [O I] 6363 |
| C III 4649 | [N II] 6548 |
| He II 4686 | H I 6563 |
| Ne IV 4720 | [N II] 6584 |
| H I 4861 | He I 6678 |
| [Fe VII] 4894 | [S II] 6716 |
| He I 4922 | [S II] 6731 |
| [Fe VII] 4943 | He I 7065 |
| [O III] 4959 | Ca B 7065 |
| [O III] 5007 | TOTL 7250 |
| [N I] 5200 | He I 7281 |
| He II 5411 | Ca II 7306 |
| N II 5495 | [OII] 7325 |
| [O I] 5577 | O IV 7713 |
| O III 5592 | C IV 7726 |
| [Ca VII] 5620 | O I 7773 |
| [Fe VII] 5721 | O I 8447 |
| [N II] 5755 | S III 9069 |

*Table 2 – Input lines in the models calculations*

| Model inputs | | | |
|---|---|---|---|
| Parameter | Range | Units | obs. |
| fc | 0 – 0.9 | | (a) |
| Shell mass | 0.01 – 20 | 10-4 Msun | |
| He | -0.67 | log(He/H) | (b) |
| C | -3.4 – -3.0 | log(C/H) | |
| N | -2.5 – -2.0 | log(N/H) | |
| O | -4.5 – -2.5 | log(O/H) | |
| clumps size | 14 – 16 | log d (cm) | (c) |
| clumps density bkg | 2 – 100 | | (d) |
| inner radius | 1 – 4 | 1016 cm | |
| outer radius | 4 – 6 | 1016 cm | |
| shell velocity | 550-300 | km/s | (e) |
| radial index | -0.9 | bkg | (f) |
| **Central Source** | | | |
| Luminosity | 35 – 37 | log erg.s-1 | |
| Temperature | 30 – 100 | 104 K | |
| Gravity | 7 – 8 | log(g) cm.s-2 | |

Table 3 – Models input parameters.
(a) – Condensed mass fraction
(b) – Abundances of He, CNO and Ne are different from solar
(c) - From line maps analysis
(d) – Density relative to background which is the r$^\alpha$ profile
(e) – Maximum and minimum expansion velocity, considering i = 40º
(f) – Index a obtained from line maps of density profile

| Line ratio | HR Del | Grid I | Grid II | Grid III |
|---|---|---|---|---|
| H I 6563/H I 4861 | 4.93 | 3.21 | 3.54 | 3.57 |
| [OIII] 5007/H I 4861 | 1.85 | 1.82 | 1.91 | 1.5 |
| He II 4686/H I 4861 | 1.01 | 0.01 | 0.06 | 1.2 |
| He I 5876/H I 4861 | 0.18 | 0.22 | 0.19 | 0.18 |
| He I 5876/He I 7065 | 15.5 | 8.91 | 5.48 | 6.2 |
| He I 6678/He I 4922 | 0.2 | 3.54 | 3.54 | 3.55 |
| [OII] 7325/H I 4861 | 0.03 | 0.001 | 0.01 | 0.02 |
| [OI] 6300/H I 4861 | 0.03 | 0 | 0 | 0 |
| [NII] 6584/H I 4861 | --- | 0.03 | 1.25 | 1.45 |
| [NII] 6548/H I 4861 | --- | 0.01 | 0.42 | 0.48 |
| [NII] 5755/H I 4861 | 0.03 | 0.006 | 0.02 | 0.02 |
| C III 4649/H I 4861 | | 0 | 0.002 | 0.005 |
| N III 4641/H 4861 | 0.17* | 0 | 0.001 | 0.001 |
| [OIII](4959+5007)/4363 | 4000** | 2000 | 2280 | 2380 |
| He I 4922/ H I 4861 | 0.08 | 0.02 | 0.02 | 0.02 |
| Temp WD (K) | | 61000 | 62000 | 65000 |
| log (L) WD erg.s-1 | | 35.5 | 35.5 | 36 |
| Mean Te Shell (K) | | 6000 | 5900 | 5800 |
| fc shell | | 0 | 0.56 | 0.7 |
| M shell (Msun) | | 1.75E-3 | 9.00E-4 | 9.00E-4 |

*Table 4 – Line ratios observed in HR Del and the models best fit.*
*\* - Both CIII and NIII lines.*
*\*\* - Upper limit of flux of [OIII]$\lambda$4363.*